\input harvmac

\Title{\vbox{\baselineskip12pt
\hbox{QMW-PH-98-17}
\hbox{gr-qc/9803095}}}
{\vbox{\centerline{Black Holes and Strings: the Polymer Link}}}

\baselineskip=12pt
\centerline {Ramzi R. Khuri\footnote{$^*$}{e-mail: 
R.R.Khuri@qmw.ac.uk. This essay received an ``Honorable Mention''
from the Gravity Research Foundation, 1998.}}
\medskip
\centerline{\sl Department of Physics}
\centerline{\sl Queen Mary and Westfield College}
\centerline{\sl Mile End Road}
\centerline{\sl London E1 4NS UK}

\bigskip
\centerline{\bf Abstract}
\medskip
\baselineskip = 20pt 
Quantum aspects of black holes represent an important testing
ground for a theory of quantum gravity. The recent success
of string theory in reproducing the Bekenstein-Hawking black hole
entropy formula provides a link between general relativity and
quantum mechanics via thermodynamics and statistical mechanics.
Here we speculate on the existence of new and unexpected links
between black holes and polymers and other soft-matter systems.

\Date{}

\def\ie{{\it i.e.,}\ }

\def\({\left (}
\def\){\right )}
\def\[{\left [}
\def\]{\right ]}

\def\buildchar#1#2#3{{\null\!
   \mathop{\vphantom{#1}\smash#1}\limits%
   ^{#2}_{#3}%
   \!\null}}

\lref\thorn{See C. B. Thorn, hep-th/9607204 and references therein; see also
O. Bergman and C. B. Thorn, Nucl. Phys. {\bf B502} (1997) 309.}

\lref\bekhawk {J. Bekenstein, Lett. Nuov. Cimento {\bf 4} (1972) 737;
Phys. Rev. {\bf D7} (1973) 2333; Phys. Rev. {\bf D9} (1974) 3292;
S. W. Hawking, Nature {\bf 248} (1974) 30; Comm. Math. 
Phys. {\bf 43} (1975) 199.}
 
\lref\GSW {M. B. Green, J. H. Schwarz and E. Witten,
{\it Superstring Theory}, Cambridge University Press, Cambridge (1987).}

\lref\prep{See M. J. Duff, R. R. Khuri and J. X. Lu, Phys. Rep.
{\bf B259} (1995) 213, M. Cvetic and D. Youm, Phys.Rev. {\bf D54}
(1996) 2612, M. Cvetic and A. A. Tseytlin, Nucl. Phys.
{\bf B477} (1996) 499 and references therein.} 

\lref\rahm{J. Rahmfeld, Phys. Lett. {\bf B372 } (1996) 198.}

\lref\pol{J. Polchinski hep-th/9611050 and references therein.}

\lref\stva {A. Strominger and C. Vafa, Phys. Lett. {\bf B379}
(1996) 99; J. Maldacena, hep-th/9607235 and references 
therein; K. Sfetsos and K. Skenderis, hep-th/9711138;
R. Arguiro. F. Englert and L. Houart, hep-th/9801053.}

\lref\corr{L. Susskind, hep-th/9309145;
G. T. Horowitz and J. Polchinski, Phys. Rev. {\bf D55}
(1997) 6189.}

\lref\self {G. T. Horowitz and J. Polchinski, Phys. Rev. {\bf D57}
(1998) 2557. See also S. Kalyana Rama, Phys. Lett. {\bf B424} (1998) 39.}

\lref\random{P. Salomonson and B. S. Skagerstam, Nucl. Phys.
{\bf B268} (1986) 349; Physica {\bf A158} (1989) 499;
D. Mitchell and N. Turok, Phys. Rev. Lett. {\bf 58} (1987) 1577;
Nucl. Phys. {\bf B294} (1987) 1138.}

\lref\polytext{See M. Doi and S. F. Edwards, {\it The Theory of 
Polymer Dynamics}, Clarendon Press, Oxford (1986) and references
therein.}

\lref\poly{S. F. Edwards and M. Muthukumar, J. Chem. Phys. {\bf 89}
(1988) 2435; S. F. Edwards and Y. Chen, J. Phys. {\bf A21}
(1988) 2963.}

\lref\callone{D. J. E. Callaway, Phys. Rev. {\bf E53} (1996) 3738.}

\lref\calltwo{D. J. E. Callaway, PROTEINS: Structure, Function
and Genetics {\bf 20} (1994) 124.}

\lref\apostol{T. M. Apostol, {\it Introduction to Analytic Number
Theory}, Springer Verlag (1976).}



The standard model of elementary particle physics has proven 
very succesful in describing three of the four fundamental 
forces of nature. In the most optimistic scenario, the standard 
model can be generalized to take the form of a grand unified theory, 
in which quantum chromodynamics, 
describing the strong force, and the electroweak theory, 
unifying the weak interaction with electromagnetism, are synthesized 
into a single theory in which all three forces have a common origin. 
The underlying framework of particle physics is   
quantum mechanics, in which the natural length 
scale associated with a particle of mass $m$ (such as an elementary 
particle) is given by the Compton wavelength $\lambda=\hbar/mc$, where 
$\hbar$ is Planck's constant divided by $2\pi$ and $c$ is the speed of 
light. Scales less than $\lambda$ are therefore unobservable within the 
context of the quantum mechanics of this particle.

Quantum mechanics, however, has so far proven unsuccessful in 
describing the fourth fundamental force, gravitation. The 
successful theory in this case is that of general relativity, which, 
however, does not lend itself to a straightforward attempt at 
quantization. The main problem in such an endeavour is that the divergences 
associated with trying to quantize gravity cannot be circumvented (or 
``renormalized") as they are for the strong, weak and electromagnetic forces. 

Among the most interesting objects predicted by general 
relativity are black holes, which represent the endpoint of gravitational 
collapse. According to relativity, an object of mass $m$ under the 
influence of only the gravitational force (\ie\ neutral with respect 
to the other three forces) will collapse into a region of spacetime bounded 
by a surface, the event horizon, beyond which signals cannot be transmitted 
to an outside observer. The event horizon for the simplest case of a 
static, spherically symmetric black hole of mass $m$
is located at a radius 
$R=2 G m/c^2$, the Schwarzschild radius, from the collapsed matter at the 
center of the sphere, where $G$ is Newton's constant. 
      
\def\buildchar#1#2#3{{\null\!
   \mathop{\vphantom{#1}\smash#1}\limits%
   ^{#2}_{#3}%
   \!\null}}

In trying to reconcile general relativity and quantum mechanics, 
a natural question to ask is whether 
they have a common domain. This would arise when an elementary particle 
exhibits features associated with gravitation, such as an event horizon. 
This may occur provided 
$\lambda \ { } 
\lower3pt\hbox{$\buildchar{\sim}{<}{}$} \ { } R$,
which implies that, even 
within the framework of quantum mechanics, 
an event horizon for an 
elementary particle may be observable. Such a condition is equivalent to
$m \ { } \lower3pt\hbox{$\buildchar{\sim}{>}{}$} \ { } m_P=
\sqrt{\hbar c/G} \sim 10^{19} GeV$, 
the Planck mass, or 
$\lambda\ { } 
\lower3pt\hbox{$\buildchar{\sim}{<}{}$} \ { } l_P
=\sqrt{\hbar G/c^3}$,
the Planck scale. 
It is in this domain that one may study a theory
that combines quantum mechanics and gravity, the so-called
{\it quantum gravity} (henceforth we use units in which
$\hbar=c=1$).

A problem, however, arises
in this comparison, because most black holes are thermal
objects, and hence cannot reasonably be identified with
pure quantum states such as elementary particles. In fact,
in accordance with the laws
 of {\it black hole thermodynamics} \bekhawk, 
black holes radiate with a (Hawking) temperature constant
over the event horizon and proportional to the surface gravity:
$T_H\sim \kappa$. Furthermore, black holes possess an entropy
$S =A/4G$, where $A$ is the area of the horizon (the area
law), and $\delta A \geq 0$ in black hole processes.
So only a black hole with zero area can correspond to
a pure state with $S=0$ such as an elementary particle, while 
a black hole with nonzero area, and therefore nonzero entropy,
corresponds to an {\it ensemble} of states. A question, then,
that can be posed of a theory of quantum gravity is the following:
since the basis of ordinary thermodynamics is (quantum) statistical
mechanics, can one recover the laws of black hole thermodynamics
by the counting of microscopic states? In particular,
can one recover the area law from a quantum mechanical entropy
arising as the logarithm of the degeneracy of quantum states?  

At the present time, string theory, the theory of one-dimensional
extended objects, is the only known reasonable
candidate theory  of quantum gravity. The divergences inherent 
in trying to quantize point-like gravity seem not to arise in
string theory. Furthermore, string theory has the potential
to unify all four fundamental forces within a common framework.
 At an intuitive level, one can see 
how point-like
divergences may be avoided in string theory by 
considering
scattering amplitudes in string theory \GSW. 
Unlike those of field theory,
the four-point amplitudes in string theory do not have 
well-defined
vertices at which the interaction can be said to take place,
hence no corresponding divergences associated with the zero size
of a particle. A simpler way of saying this is that the finite 
size of 
the string smooths out the divergence of the point particle.

{}For the purpose of understanding black hole thermodynamics,
an important feature of string theory is that classical 
solutions \prep\ may be easily constructed as composites  
of single-charged fundamental constituents. Identifying
these constituents with states in string theory, one can
compare the Bekenstein-Hawking entropy obtained from the
area of the classical solution to the quantum-mechanical
microcanonical counting of ensembles of states \stva.
For example, the extremal
Reissner-Nordstr\"om charged black hole solution
of Einstein-Maxwell theory arises in string theory
as the composite of four charges, 
$N_1$, $N_2$, $N_3$ and $N_4$, normalized to correspond to
number operators in string theory.
The area law then yields a Bekenstein-Hawking
entropy $S_{BH}=2\pi \sqrt{N_1 N_2 N_3 N_4}$. The 
counting of the degeneracy of the states forming this black hole
leads to the same quantity $S_{QM}=\ln d(N_i)=S_{BH}$. 
Even in the black hole picture, this result can be seen to arise
from the number of ways in which the various constituents combine.
Following \rahm, one can write four-centered solutions
each with charge $N_i$ of a given species. A black hole with
nonzero area is formed when all charges are brought 
together to the same point. The precise partition function \apostol\
yielding the correct degeneracy $d(N_i)=exp(S_{BH})$ is obtained
provided both bosonic and fermionic excitations of a supersymmetric
string-like object along various dimensions are taken into account.

The recovery of the area law in a wide variety of
contexts in string theory suggests that we have accounted
for the microscopic degrees of freedom of the black hole.
However, the ensemble of string states on the one hand
and the black hole on the other represent two very different 
objects, so we must try to understand
the correspondence between them \corr. For
simplicity, let us consider the case of a long,
self-gravitating string in $D=4$ dimensions \self.
At level $N$, a free string has
mass $M\sim \sqrt{N}/l_s$, size $L \sim N^{1/4} l_s$
and entropy $S\sim\sqrt{N}$, 
where $l_s$ is the string scale. This picture is valid
provided the string coupling $g<<1$, where $g$ is
related to Newton's constant $G$ via
$G \sim g^2 \l_s^2$. This picture represents a random walk \random\
with $n=\sqrt{N}$ steps, each a single string ``bit'' of length $l_s$
\thorn.

Let us now slowly increase the coupling $g$.
As shown in \self, gravitational effects start becoming
strong at $g_0 \sim N^{-3/8}=n^{-3/4}$, after which the string 
collapses until it reachers the size of the string scale $l_s$. 
At the critical coupling
$g_c \sim N^{-1/4}=n^{-1/2}$, the Schwarzschild radius
$R=2GM$ of a black hole with the same mass becomes
of the order of the string scale, and one can sensibly
start thinking of the string as a black hole. At this point,
too, the entropies match: $S_{BH} \sim R^2/G=1/g_c^2=\sqrt{N}=n$.
For $g>g_c$, the black hole picture prevails.
In the intermediate range $g_0 < g < g_c$, the size of the
string state was shown using a thermal scalar field
theory in \self\ to be 
\eqn\inter{L\sim {l_s\over g^2 N^{1/2}}={l_s\over g^2 n},} 
which smoothly interpolates between the random walk size and the 
string scale. Note that for $n$ large, the coupling is
small throughout the ranges we are considering.
This is an interesting result with a specific prediction
for the coupling dependence of the size of the string as it
collapses into a black hole. A natural question to ask is whether
this sort of result also arises in analogous physical systems already 
considered. Since random walks with interactions arise in polymer
physics \refs{\polytext,\thorn}, the relation \inter\ should also hold
for a self-attracting polymer chain.

We start with a random walk with $n$ steps each
of size $a$, so that the size of the polymer is initially given
by $L_0=\sqrt{n} a$. Suppose we place
the polymer in a medium of scatterers of
number density $\rho$ and (dimensionless) potential strength
$u$. Then the size of the polymer was shown to be \poly\
\eqn\sizeone{L^2=x^{-2} \left(1-\exp(-nx^2a^2)\right),}
where $x=u\rho a^2$
can be thought of as an effective scattering cross section.

To compare with a self-gravitating string with $a=l_s$, 
the scatterers
are taken to coincide with the positions of the string bits
themselves. For large $n$ and in a mean-field approximation,
the number density of $n$ bits in a volume $L_0^3$ is given
by
\eqn\density{\rho={n\over (n^{3/2}l_s^3)}=n^{-1/2} l_s^{-3}.}
For $g$ small, the leading order interaction potential is 
given by
\eqn\potential{{u\over l_s}\sim \sum_{i,j} {g^2\over 
|\vec r_i - \vec r_j|} \sim {g^2 n^2\over L_0}
={g^2 n^{3/2}\over l_s},}
where $\vec r_i$ is the position of the $ith$ link.
It follows that $x\sim ng^2/l_s$, so that  
\eqn\sizetwo{L^2=l_s^2 n^{-2} g^{-4}
\left( 1- \exp(-n^3 g^4) \right).} 
For $g < g_0=n^{-3/4}$, $L^2 \simeq n l_s^2$ which is 
the random walk, corresponding to the free string.
As in the string case, a transition occurs at $g\sim g_0$. 
As $g$ is increased past $g_0$, the size quickly shrinks to
$L^2 \simeq l_s^2/n^2 g^4=l_s^2/g^2 N$,
as in \inter. This kind of relation holds\foot{Once the 
self-interaction
of the polymer becomes strong, the simple result \sizetwo\ 
is no longer exact and a more precise computation is required.
Nevertheless, it is clear that one obtains a smooth
transition from the random walk to the Schwarzschild radius
via a nonperturbative coupling dependence, so that even if
\inter\ is not exactly recovered, it remains a good approximation
for the collapse of the polymer.}
until $g\sim g_c \sim n^{-1/2}$,
when $L\sim R$, the Schwarzschild radius of the polymer, 
and the black hole picture dominates.  

This connection between black holes, strings and polymers is
very interesting and merits further investigation. Similar
links with other soft-matter systems have also been noted
in \callone, where the area law was recovered for the
case of a liquid field theory and where it was argued
that the area law contributions to the free energy are
primarily responsible for liquid surface tension.
The speculation was also made that the area law arises 
in the context of protein folding.

Connections between physical and biological systems are 
always exciting. The cases discussed above are especially so since
quantum gravity is generally considered 
too remote to have relevance to other areas of physics, much less
other fields of science. In particular, the fascinating possibility
arises that mathematical techniques used to study black holes
can be useful in understanding biological questions, such as 
protein dynamics, while methods of polymers physics can
potentially shed light on quantum gravity.

\bigskip
I would like to thank Rob Myers, Amanda Peet and Gary Horowitz 
for helpful discussions. I would also like to thank David Gross,
the ITP and the Department of Physics at UCSB for their 
hospitality and where part of this work was done. 
Research supported
by a PPARC Advanced Fellowship. This research was also supported 
in part by the National Science Foundation under Grant No. PHY94-07194.

\listrefs
\end